%% file: main_upload.tex
\documentclass[11pt,a4paper,reqno]{amsart}

\usepackage{siunitx}
\usepackage{upgreek}
\usepackage{natbib}
\usepackage{tabularx}
\usepackage{graphicx} 
\usepackage[headings]{fullpage}
\usepackage{amsaddr}

\def\SymbReg{\textsuperscript{\textregistered}}
\newcolumntype{Y}{>{\centering\arraybackslash}X}

\begin{document}
\title[Characterization of the 3D microstructure of Ibuprofen tablets]{Characterization of the 3D microstructure of Ibuprofen tablets by means of  synchrotron tomography}

\author[Neumann, Cabiscol, Mark\"{o}tter, Osenberg, Manke, Finke, Schmidt]{Matthias~Neumann$^1$, Ramon~Cabiscol$^{2,3}$, Markus~Osenberg$^4$, Henning~Mark\"{o}tter$^5$, Ingo~Manke$^5$, Jan-Henrik~Finke$^{2,3}$, Volker~Schmidt$^1$}

\address{$^1$Institute of Stochastics, Ulm University, Helmholtzstr. 18, 89069~Ulm, Germany}
\address{$^2$Institute for Particle Technology (iPAT), TU Braunschweig, Volkmaroder Str. 5, 38104~Braunschweig, Germany}
\address{$^3$Center of Pharmaceutical Engineering (PVZ), Franz-Liszt-Str. 35-A, 38106~Braunschweig, Germany}
\address{$^4$Department of Materials Science and Technology, Technische Universität Berlin, Hardenbergstr. 36, 10623~Berlin, Germany}
\address{$^5$Institute of Applied Materials, Helmholtz-Zentrum Berlin, Hahn-Meitner-Platz 1, 14109~Berlin, Germany}

\keywords{3D imaging, Ibuprofen tablet, microstructure characterization, statistical learning, image trinarization, watershed algorithm}

\begin{abstract}
We present a methodology to characterize the 3D microstructure of Ibuprofen tablets, which strongly influences the strength and solubility kinetics of the final formulation, by means of synchrotron tomography. A physically coherent trinarization for greyscale images of Ibuprofen tablets consisting of the three phases microcrystalline cellulose, Ibuprofen and pores is presented. For this purpose, a hybrid approach is developed combining a trinarization by means of statistical learning with a trinarization based on a watershed algorithm. This hybrid approach allows us to compute microstructure characteristics of tablets using image analysis. A comparison with experimental results shows that there is a significant amount of pores which is below the resolution limit and results from image analysis let us conjecture that these pores make up the major part of the surface between pores and solid. Furthermore, we compute microstructure characteristics, which are experimentally not accessible such as local percolation probabilities and chord length distribution functions. Both characteristics are meaningful in order to quantify the influence of tablet compaction on its microstructure. The presented approach can be used to get better insights into the relationship between production parameters and microstructure characteristics based on 3D image data of Ibuprofen tablets manufactured under different conditions.
\end{abstract}

\maketitle

\section{Introduction}\label{sec:intro}
\input{intro}

\section{Materials and experimental methods}\label{sec:materials}
\input{materials}

\section{Experimental results}\label{sec:exp_results}
\input{exp_results}

\section{Imaging}\label{sec:imaging}
\input{imaging}

\section{Trinarization of grey-scale images}\label{sec:tri}
\input{tri}

\section{Statistical microstructure characterization}\label{sec:image_analysis}
\input{image_analysis}

\section{Conclusions}\label{sec:conclusions}
\input{conclusions}

\bibliographystyle{rss}
\bibliography{C:/Users/matthias/Bib/Literatur}

\newpage
\section*{Figures}
\input{figures0}

\newpage
\section*{Tables}
\input{tables}

\end{document}

%% file: intro.tex
In the past years, the so-called ``Quality by design'' principle has been targeted as the key strategy for the development and manufacturing of novel pharmaceutical products. This new method disrupts with the traditional ``trial-and-error'' and ``quality-by-end-product'' paths and seeks to foresee the performance of the final product from the understanding of process and component attributes. Direct compaction, a critical stage in the production of pharmaceutical tablets, is a main example of ``trial-and-error'' operation. The absence of defects throughout the operation (delamination, capping or chipping), a good powder flowability and an adequate breakage resistance constitute the product conformity criteria~\citep{zhou.2010}.

When it comes to a commercial multi-component tablet, the interplay of each constituent of the powder blend (excipients, lubricants, active ingredients, among others) affects drastically the strength and solubility kinetics of the final formulation. Many factors of the blends influence the properties of the final tablet such as porosity, shape, surface area and size~\citep{sebhatu.1999, olsson.2001, poquillon.2002}. Several empirical approaches have been reported in order to model the tensile strength from blend properties. \cite{chan.1983} provided a model considering the effects of particle size and composition of binary mixtures. Based on percolation theory,~\cite{kuentz.1998} developed a model for the tensile strength of binary formulations, assuming that a tablet can only be produced with a solid fraction larger than some critical threshold, which is required to build a percolating system in the tablet.

However, these empirical relationships between properties of single component constituents of binary tablets and tensile strength do not capture microstructure characteristics like connectivity and surface area of each of the constituents, which also influence the strength of the tablet. The spread of visualization techniques resulting from computer aided reconstructions has opened a whole range of possibilities for a detailed microstructure analysis of tablets. Several imaging techniques have been applied towards this goal, including synchrotron tomography, terahertz pulse, Raman and NIR spectroscopy. For an overview, the reader is referred to~\citep{muellertz.2016} and the references therein. In the present study, a methodological approach to investigate binary tablets based on synchrotron tomography with a voxel size of 0.44 $\upmu$m is presented and demonstrated exemplarily for bicomponent tablets consisting of microcrystalline cellulose (MCC) and Ibuprofen (API) and working as follows. Water accesses the inner parts of the tablet via the pores and dissolves the MCC. Then, API is step by step given to the body. Since the underlying microstructure of the tablet influences the flow of water through the pores as well as the solubility kinetics, it influences the speed and amount of API given to the body. While the microstructure of individual MCC particles has recently been investigated based on synchrotron tomography with a voxel size of 0.65 $\upmu$m~\citep{fang.2017}, our focus goes beyond and provides a characterization of all three phases in the tablet, i.e. the constituents MCC and API as well as pore space.

For this purpose, it is necessary to suitably trinarize the 3D images, i.e. each voxel has to be labeled as MCC, API or pore. Classical algorithms from morphological image analysis as reviewed in \cite{schluter.2014} and algorithms based on statistical learning~\citep{james.2013} do not lead to a sufficiently good trinarization. More precisely, they are not able to classify the pore space adequately. In order to overcome this limitation a new method to trinarize the greyscale images obtained by synchrotron tomography is developed. We propose a hybrid approach combining a watershed-based trinarization going back to ideas presented in~\cite{meyer.1990} with a trinarization using a random forest algorithm, a tool from statistical learning. Another combination of tools from mathematical morphology with statistical learning has recently been used to improve the quality of particle-wise segmentation from 3D image data~\citep{furat.2018}. In the present paper, the hybrid approach leads to a trinarization allowing for the computation of structural characteristics, which are only accessible via image analysis such as, e.g., local percolation probabilities or chord length distribution functions. These characteristics can be used for the quantification of anisotropy effects in a microstructure resulting from a uniaxial compaction of the tablet. The values of porosity and specific surface area obtained from image analysis are compared with the corresponding experimentally obtained values. The methodology considered in the present paper permits us to extend the characterization of the 3D microstructure of Ibuprofen tablets by image analysis, which is a first step towards the goal to relate microstructure characteristics with product conformity criteria like powder flow and breakage resistance. 

The paper is organized as follows. Materials and experimental methods are described in Section~\ref{sec:materials}, whereas the results of experimental measurements are given in Section~\ref{sec:exp_results}. After the description of 3D imaging in Section~\ref{sec:imaging}, we present a hybrid approach for the algorithmic trinarization of 3D images in Section~\ref{sec:tri}. The algorithm is tested for different cutouts of one large 3D image of the microstructure of Ibuprofen tablets which allows for a statistical characterization of the microstructure by image analysis given in Section~\ref{sec:image_analysis}. Section~\ref{sec:conclusions} concludes the work.

%% file: materials.tex
\subsection{Materials}

Binary tablets of microcrystalline cellulose (MCC) (Vivapur\SymbReg 12, JRS Pharma, Germany) and Ibuprofen Gracel are used to produce the three-phase tablets considered in this paper. In order to narrow the particle size distribution (PSD) of Ibuprofen, the fraction of Ibuprofen above $180~\upmu\mathrm{m}$ remaining after manual sieving is removed before admixing.

Due to the poor processability of this mixture, each batch is internally lubricated by blending of Magnesium stearate (MgSt) (Magnesia 4264; Magnesia GmbH, Germany). Blends are homogenized in a powder mixer ERWEKA AR-403/-S (ERWEKA GmbH, Germany) with a 3.5 l cubic container for 3 minutes at 16 rpm before tableting. Two different formulations were considered for this analysis:\newline a) 94.05 \% (w/w) MCC; 4.95 \% (w/w) API and 1.00 \% (w/w) MgSt, b) 79.20 \% (w/w) MCC; 19.80 \% (w/w) API and 1.00 \% (w/w) MgSt.

General powder characteristics, such as particle size distribution (PSD) and true density are determined. Powder particle size distribution is determined via dry dispersion by an airflow injector (Mastersizer 300, Malvern Instruments Ltd, United Kingdom) using a differential pressure of 2~bar.

Density of both powders is required in order to determine the evolution of internal tablet porosity with the compaction pressure. Skeletal or true density of primary particles is determined by means of helium pycnometry (ULTRAPYC 1200e, Quantachrome GmbH, Germany). Samples were stored at $20 ^\circ\mathrm{C}$ and 45 \% RH during the 24~h prior to the analysis. Measured values are extracted after 10 runs and assuring a typical deviation lower than 0.005.

Blends are compacted with the compaction simulator STYL'One Evolution (Medel Pharm S.A.S., France). Standard EURO B die and punches were set up in order to produce cylindrical tablets of 11.28 mm in diameter. The compaction sequence comprises the filling of the die by gravity up to a height of 10 mm with the blend of interest (previously stored at $20 ^\circ\mathrm{C}$ and 45~\% RH for 24~h) and the symmetrical movement of the punches at a constant speed of 20.6~mm/s until the target pressure was achieved. In the current study compaction pressures running from 45~MPa up to 188~MPa were analyzed.

\subsection{Experimental methods}

Internal surface area of tablets was determined by nitrogen sorption at the ASAP 2460 (Micromeritics Instrument Corp., USA). Sample conditioning proceeds as follows: after compaction, cylindrical tablets are cut up in quarters and inserted into the degassing units where they were treated for 24~h at room temperature at vacuum conditions in order to remove physisorpted compounds. Then, a conditioning interval of 500~s precedes the sorption of nitrogen at relative nitrogen pressure range $P/P_0$ from 0.10 to 0.30 and an absolute temperature of $-196^{\circ}\mathrm{C}$. Samples were measured in triplicate. Finally, the best linear fit was obtained for the BET model. 

Pore size analysis experiments were executed using a mercury intrusion porosimeter (MIP) PoreMaster 60 (Quantachrome GmbH, Germany). Pressures ranging from 1 to 60,000 PSI were applied in a high pressure station. The pressure was exerted onto the sample using a penetrometer made of glass as specimen container. A penetrometer with a stem volume of $0.5~\mathrm{cm}^3$ and a sample container of $3.8~\mathrm{cm}$ length is used. 

One of the main assumptions of this technique is cylindrical shape pore configuration. Based on that, the MIP pore size distribution can be determined using a modified Young-Laplace equation, referred mainly as Washburn equation~\citep{washburn.1921}:
\begin{equation}
\Delta P = \gamma \left(\frac{1}{r_1} + \frac{1}{r_2} \right)= \frac{2\gamma \cos \theta}{r_{\mathrm{pore}}},
\end{equation}
where $\Delta P$ is the pressure difference, $r_1$  and $r_2$ describe the curvature of the interface, $r_\mathrm{pore}$ the pore size using the surface tension of mercury $\gamma$ and the contact angle $\theta$ between the solid and mercury. A basic assumption for this relation is a constant surface tension $\gamma =0.485~\mathrm{Nm}^{-1}$ and contact angle $\theta=140^{\circ}$ of the intruded mercury and the substrate. According to the intrumental pressure range, a pore size span from 1.80 nm to 108 $\upmu$m should be accessible with this setup. The MIP pore size distribution is denoted by $Q_3$ in the following.

%% file: exp_results.tex
The density of the considered tablets experimentally determined by helium pycnometry is 1279 kg/m$^3$ and the specific surface area determined by BET measurements is between 0.78~$\upmu\mathrm{m}^{-1}$ and 0.81~$\upmu\mathrm{m}$. In Section~\ref{sec:image_analysis} these results are compared with the results of image analysis. The cumulative volumetric pore size distribution is presented in Figure~\ref{fig:Q3}. The results show that roughly 25~\% of the pores have a diameter of less than $0.438~\upmu\mathrm{m}$ which is the voxel size of the 3D images and that most of the pores have a diameter which is lower than the resolution limit (about $2~\upmu\mathrm{m}$ after filtering, see Section~\ref{subsec:synchrotron}). Note that the pores below the resolution of image data are inaccessible for image analysis. 

The 10\%-,50\%- and 90\%-quantiles of the volumetric/mass particle size distribution (PSD), denoted by $X_{10}$, $X_{50}$ and $X_{90}$, and the true density of MCC and Ibuprofen are summarized in Table~\ref{tab:exp_results}. In constrast to the pore size, the values of $X_{10}$ of both powders are some orders of magnitude higher than the resolution of image data. Therefore primary particles should be satisfactorily detected by synchrotron tomography.

%% file: imaging.tex
\subsection{Sample preparation}

First measurements have shown that small empty pores lead to imaging artifacts arising due to refraction at the gas/material interface. This prevents successful classification/trinarization of the tablets material composition. Therefore, the pores are filled by a contrast medium, which reduces the refraction at the pore/material interface. This is achieved by cutting a piece of the Ibuprofen tablet and fixing it into a polyimide tube with an inner diameter of 1.6~mm, which is then filled up with a hydrocarbon based contrast medium and sealed on both tube ends. The tube with the sample therein is then imaged via synchrotron tomography.

\subsection{Synchrotron tomography}\label{subsec:synchrotron}

The synchrotron tomography measurement is conducted at the BAMline at the electron storage ring Bessy II in Berlin, Germany~\citep{goerner.2001}. For optimized contrast of the light elements in the sample a beam energy of 9.8 keV is chosen with a double multilayer monochromator. After transmitting the sample, the X-ray beam is converted with a 60~$\upmu\mathrm{m}$ thick $\mathrm{CdWO}_4$ scintillator into visible light, which is then imaged onto the CCD chip of a PCO.4000 camera (PCO AG). The used optical setup yields a resolution of $4008 \times 2672$ pixels covering an area of $1.75 \times 1.17~\mathrm{mm}^2$, which corresponds to a pixel size of $0.438~\upmu\mathrm{m}$. For the tomographic scan 2200 projections covering an angular range of $180^{\circ}$ are collected and used for the reconstruction via filtered back projection~\citep{jaehne.2005}. Each projection is exposed for 3~s. Together with flat field images for image normalization this results in a total scan time of approximately 2~h. A phase retrieval algorithm, according to \cite{paganin.2002}, is applied on the projections in order to emphasize the contrast evoked by the phase shift in the material. After the application of the phase retrieval algorithm, the resolution limit is at $\approx 2~\upmu\mathrm{m}$.

%% file: tri.tex
In order to investigate the microstructure of an Ibuprofen tablet by statistical image analysis, the grey-scale images have to be trinarized. Figure~\ref{fig:tri2D}a shows that 3D imaging described in Section~\ref{sec:imaging} leads to a good contrast between pores, MCC and API. The darkest greyscale values correspond to the pore space, the medium values to API and the brighter values to MCC. Despite of the good contrast, algorithmic trinarization encounters two major challenges. On the one hand, there are voxels within MCC, the greyscale values of which are in the same range as the ones of voxels belonging clearly to API. Moreover, the greyscale values of thin pores, which are located at the boundary between different particles of MCC or which are located within a particle of MCC, are similar to the greyscale values of API. From a physical point of view, it is not reasonable to rely only on thresholding of greyscale values. 

The trinarization proposed in this paper is based on a random forest algorithm, which meets the above mentioned challenges (Figure~\ref{fig:tri2D}). In a second step, the results of this trinarization are improved by a combination with a trinarization based on the watershed algorithm (Figure~\ref{fig:tri2D}c). This leads to a hybrid approach used for the final trinarization (Figure~\ref{fig:tri2D}d), which takes benefit from the advantages of both trinarization algorithms.

\subsection{Trinarization by a random forest algorithm}

A random forest is an algorithm for classification from statistical learning based on decision trees.  In contrast to classification by a single decision tree, a random forest is built by a large number of randomized decision trees. For a detailed description of random forests we refer to~\cite{james.2013}. In the present paper, a random forest algorithm is used for the trinarization of greyscale images, which can be considered as a classification problem. Each voxel has either to be classified as pore, API or MCC. 

The random forest has to be trained, which means that, roughly speaking, the algorithm has to learn how to trinarize a given greyscale image. For this purpose, a 2D slice of the 3D image is considered, in which $N$ voxels are manually trinarized by visual inspection. Thereby, a trinarization mask for the selected 2D slice is obtained. The same 2D slice is then filtered in $M$ different ways, such that we result in an $N\times(M+2)$-dimensional matrix. The rows of this matrix indicate the pixels, which are considered during the manual trinarization. In each row the original greyscale value of the pixel, its greyscale values after the application of each of the $M$ filters as well as its manually determined class label are stored. The matrix represents the training data for the random forest. 

In order to perform training and feature evaluation, we use Ilastik~\citep{sommer.2011} with the parallelized random forest implemented in the computer vision library VIGRA. Finally, the $M$ filters are applied to all slices of the 3D images, which allows for a trinarization of all voxels in the 3D image based on the trained random forest. The random forest algorithm, the result of which is visualized in Figure~\ref{fig:tri2D}b, works satisfactorily for trinarizing the considered image data. In particular, thin pores (e.g. green circles in Figure~\ref{fig:tri2D}b) and darker greyscale values within MCC (e.g. blue circle in Figure~\ref{fig:tri2D}) are properly resolved. For example, the long and thin pore in the upper left part of the 2D slice in Figure~\ref{fig:tri2D}b is recognized correctly. Furthermore, the random forest algorithm does not lead to misclassification within particles of the MCC, even if smaller greyscale values may suggest an occurrence of API.    
However, comparing the result of the random forest algorithm with the original greyscale image one can observe that pores are detected between API and MCC, although there is no indication for pores, neither by greyscale values nor by physical reasons (e.g. red circles in Figure~\ref{fig:tri2D}b). This effect occurs because the algorithm is trained to detect pores at the boundary of MCC. Moreover, unrealistically tiny connected components of all three phases are created using the random forest algorithm. Both effects can be corrected by using a hybrid approach for trinarization, i.e. by a combination of the random forest algorithm with a trinarization based on the watershed algorithm, see Section~\ref{subsec:hybrid}.

\subsection{Watershed-based trinarization}

The watershed algorithm is a method to partition an image into different regions, so-called watershed basins, see e.g.~\cite{roerdink.2000}. Consider any 3D image as a landscape, where its greyscale values as the local altitudes of the landscape. Then, the regions generated by the watershed algorithm can be interpreted as valleys, where their boundaries are given by the watershed lines of the landscape. Typically used for the extraction of individual particles from image data, the watershed algorithm has also been applied to classify different phases in multi-phase materials~\citep{schluter.2014}. In the present paper, we propose a new method how to use the watershed algorithm to trinarize greyscale images of three-phase materials. 

The principal idea of the trinarization is the following. As a starting point, the pore space is classified by global thresholding, before the solid phase, i.e. the union set of MCC and API, is partitioned by a special type of watershed algorithm. According to its average greyscale value, each watershed basin is finally assigned to one of the three phases. Note that averaging over all greyscale values within a watershed basin has been used by~\cite{frucci.2013} for filtering images. 

For a more detailed description of the watershed-based trinarization, we denote the set of voxels by $W\subset \mathbb{Z}^3$. The greyscale image is denoted by $I$ and for each voxel $w\in W$, we denote its corresponding greyscale value by $I(w)$. To distinguish the pore space from the solid phases we use global thresholding, where the threshold value $t_1$ is determined by visual inspection. This leads to a numerical value of $t_1=30739$. Note that we deal with 16-bit images, i.e., all greyscale values are between 0 and 65535. The determined pore space for a cutout of a 2D slice is visualized in Figure~\ref{fig:watershed2D}b.

In the next step, a further image $J$ is defined on the basis of which the watershed basins are computed. The greyscale values of $J$ are low at voxels, for which it is clear whether they belong to MCC or API. Considering the image $J$ as a landscape, as the watershed algorithm does, the uncertainty of classification is low at voxels with a low altitude. In turn, the uncertainty of the classification is high for voxels located close to the watershed lines. Thus, the algorithm is constructed such that phase transition only occurs at watershed lines. 

Formally, $J$ is defined as follows. Applying the iterative algorithm introduced by~\cite{ridler.1978} to all voxels, not classified as pores before, we obtain a threshold between MCC and API, denoted by $t_2$. The voxels of greyscale values differing much from $t_2$ are considered as voxels which can be easily classified. Thus, greyscale values of $J$ are defined by 
\begin{equation}
J(x)=\left\{\begin{array}{ll} - ( I(w) - t_1)^2, & \text{if~} I(w) \geq t_1,  \\
0, & \text{else.}\end{array}\right.
\end{equation}
After smoothing $J$ by a minimum filter with a radius of 2 voxels $(\approx 0.876~\upmu\mathrm{m})$, the watershed algorithm is applied on $J$, see Figure~\ref{fig:watershed2D}c. For this purpose, we use the algorithm which was introduced by~\cite{meyer.1994}. For an overview on watershed algorithms, we refer to~\cite{beare.2006}. 

The watershed basins are assigned to MCC or API by global thresholding with respect to the average greyscale values of the basins. Here $t_2$ is used as global threshold. The voxels, located on the watershed boundaries are finally classified by an application of a maximum filter. The resulting trinarization contains unrealistically many small connected components of API within the MCC. Therefore, in a post-processing step, all API clusters with less than 300000 voxels $(\approx 25200~\upmu\mathrm{m}^3)$ are removed and assigned to MCC. Note that whenever cluster analysis is performed in the present paper, the algorithm of~\cite{hoshen.1976} is used.

The result of the watershed-based trinarization described above is visualized in Figure~\ref{fig:tri2D}c. Regarding the detection of small pores at the boundary of MCC and the representation of MCC itself, the random forest algorithm leads to a better trinarization (e.g. red circles in Figure~\ref{fig:tri2D}c). However, the watershed algorithm does not detect unrealistic pores between MCC and API as the random forest algorithm does (e.g. green circles in Figure~\ref{fig:tri2D}c). Thus, we can use the results obtained by the watershed-based trinarization to improve the trinarization by the random forest algorithm.

\subsection{Hybrid approach}\label{subsec:hybrid}
The trinarization obtained by the random forest algorithm, denoted by $T_1$, is the basis of the hybrid approach. It is just modified using information from the watershed-based trinarization, denoted by $T_2$. In a first step, tiny connected components of all three phases are removed from $T_1$. All connected components of the pore space with less than 1000 voxels~$({\approx84}~\upmu\mathrm{m}^3)$ are assigned to API. After that, all connected components of API with less than 8000 voxels~$({\approx670}~\upmu\mathrm{m}^3)$ are assigned to the pore phase. Finally all connected components of MCC with less than 8000 voxels~$({\approx670}~\upmu\mathrm{m}^3)$ are assigned to API. Note that by this procedure tiny pores, which are completely surrounded by MCC, are not removed. The existence of those pores is not in contradiction with the prior knowledge about the material from a physical perspective.  

Finally, all unrealistic pores between MCC and API detected in $T_1$ are removed. For this purpose, we assign each voxel $w \in W$, which belongs to the pore phase in $T_1$, to API if the following two conditions are fulfilled: 1) The Euclidean distance from $w$ to the closest pore voxel in $T_2$ is larger than 20 voxel~$(\approx 8.76~\upmu\mathrm{m})$. 2) The Euclidean distance to the closest API voxel in $T_1$ is smaller than 20 voxel~$(\approx 8.76~\upmu\mathrm{m})$. While Condition 1 uses $T_2$ to check if the pore voxel is wrongly detected in $T_1$, Condition 2 ensures that the considered pore voxel is close to API. Condition 2 is necessary because pores within MCC, which are not detected by the watershed-based trinarization, should not be removed.

The results obtained by the hybrid approach, see Figure~\ref{fig:tri2D}d, show that the required challenges can be met by combining the random forest algorithm with the watershed-based trinarization. Note that the tiny connected components, which can be observed in Figure~\ref{fig:tri2D}d are connected in 3D. A 3D visualization of MCC as well as API determined by the hybrid approach is given in Figure~\ref{fig:3Dvis}.

%% file: image_analysis.tex
Trinarization is performed on eight non-overlapping cubic cutouts with a side length of $306.6~\upmu\mathrm{m}$ of the entire 3D image. The hybrid approach introduced in Section~\ref{subsec:hybrid} is used for this segmentation. The trinarized cutouts are statistically analyzed using methods of spatial statistics, see e.g.~\cite{chiu.2013}. In Section~\ref{subsec:volume_surface} the results of the statistical analysis of trinarized images are compared with the results obtained by the experimental characterization of tablets described in Section~\ref{sec:exp_results}. 

\subsection{Volume fraction and surface area}\label{subsec:volume_surface}

We consider the volume fractions and the specific surface areas of the three phases. The volume fractions of MCC and API and their specific surface areas estimated from image data are visualized in Figure~\ref{fig:vol_surf}. Note that for a given porosity/density, the volume fraction of MCC is a function of the volume fraction of API. The graphs of these function are visualized for certain fixed values of density and porosity in Figure~\ref{fig:vol_surf}a. One can observe that volume fractions of MCC are between 0.7 and 0.77, while the volume fractions of API are between 0.21 and 0.28. Porosity varies between 0.025 and 0.075. A density between $1350~ \mathrm{kg}/\mathrm{m}^3$ and $1450~\mathrm{kg}/\mathrm{m}^3$ is computed using the volume fractions of phases. These values are larger than the experimentally determined density of $1279~\mathrm{kg}/\mathrm{m}^3$. This overestimation is attributed to pores, the size of which is below the voxel size of $0.44~\upmu\mathrm{m}$ and which are thus not visible in the 3D images. Moreover, note that the resolution of 3D images is about $2~\upmu\mathrm{m}$, which does not allow for an exact determination of porosity via image analysis. The existence of such small pores is proven by the experimental results of mercury intrusion porosimetry, see Figure~\ref{fig:Q3}.

Moreover, the specific surface area, which is defined as the ratio of surface area and the volume fraction of the solid phase, is estimated from image data using the method given by~\cite{ohser.2009}. The values are higher for MCC than for API and pores since MCC has the highest volume fraction of the three phases. The specific surface area of pores obtained from image analysis is much lower than the experimental values, which are between $0.78~\upmu\mathrm{m}^{-1}$ and $0.81~\upmu\mathrm{m}^{-1}$. Thus we conjecture that most parts of the surface area of pores come from the thin pores which are below the resolution limit.

\subsection{Further microstructure characteristics}

Besides estimating the surface area of each phase, the surface area of the interfaces, between pores and MCC, between pores and API, and between API and MCC can be computed. The surface areas of interfaces per unit cube are denoted by $I(\mathrm{pore},\mathrm{MCC}), I(\mathrm{pore},\mathrm{API}),$ and $I(\mathrm{API},\mathrm{MCC})$, respectively. Note that
\begin{equation}
I(\mathrm{pore},\mathrm{MCC}) = \frac{1}{2} \,(S_\mathrm{pore}+S_\mathrm{MCC}-S_\mathrm{API}),
\end{equation} 
where $S_i$ denotes the surface area of phase $i$ per unit cube. The surface areas of the other interfaces can be computed analogously. This allows us to determine the proportion of pairwise interfaces in the complete surface area, see Figure~\ref{fig:interface}a. Depending on the cutout, the proportion of $I(\mathrm{pore},\mathrm{MCC})$ and $I(\mathrm{API},\mathrm{MCC})$ varies strongly for the eight samples in the range of $0.2-0.45$ and $0.45-0.7$, respectively. This variation can be explained by the variation of porosity, see Section~\ref{subsec:volume_surface}. Further underpinning of this explanation is given by the fact that both, $I(\mathrm{pore},\mathrm{MCC})$ and $I(\mathrm{pore},\mathrm{API})$ strongly correlate with porosity, see Figure~\ref{fig:interface}b.

Next, local percolation probabilities~\citep{hilfer.1991} are computed to quantify connectivity properties of the three phases API, MCC and pores, see Figure~\ref{fig:percolation}. For this purpose, we divide each cubic cutout in sub-cubes with an edge length of $22~\upmu\mathrm{m}$. For each sub-cube and each phase, we compute the volume fraction of the phase and check if it is percolating in $x$-, $y$-, and $z$-direction. We say that a phase is percolating within a sub-cube $[0,22~\upmu\mathrm{m}]^3$ in, e.g., $x$-direction if there exists a path (with respect to the 26-neighborhood, i.e. each voxel $w \in W$ is connected to all other voxels which share at least one vertex with $w$~\citep{ohser.2009}) from the bottom $\lbrace 0 \rbrace \times [0,22~\upmu\mathrm{m}]^2$ to the top $\lbrace 22~\upmu\mathrm{m} \rbrace \times[0,22~\upmu\mathrm{m}]^2$ within the considered phase. Then, the percolation probability $P(v)$ for a given volume fraction $v$ is estimated by a Nadaraya-Watson estimator~\citep{nadaraya.1964} with a Gaussian kernel and a manually chosen bandwidth of $h=0.3$. In Figure~\ref{fig:percolation} histograms representing the frequency of volume fractions of the sub-cubes are given for each of the three phases. Since we do not observe sub-cubes with a porosity of more than $0.5$, the percolation probabilities for higher porosities cannot be considered as accurate estimates. In Figure~\ref{fig:percolation}a, the black dotted line shows the upper volume fraction limit, where the estimation of percolation probabilities is no longer meaningful. Formally, we consider an estimate of $P(v)$ as meaningful if the denominator in the Nadaraya-Watson estimator is larger than $100$. For API and MCC, we observe sufficiently many sub-cubes for the whole range of volume fractions in order to estimate the percolation probabilities. One can observe for all three phases that the percolation probabilities increase stronger with increasing volume fractions in $x$- and $y$-direction than in $z$-direction. This result is reasonable as the tablet is compacted in $z$-direction. Thus, the pores, as well as API and MCC are more elongated in the $xy$-plane leading, in turn, to better connectivity of all three phases in $x$- and $y$-direction than in $z$-direction. The computation of local percolation probabilities can be considered as a quantification of the influence of compaction on the tablet microstructure.

Another relevant descriptor of the microstructure is the so-called chord length distribution function. Note that the chord length distribution in a certain direction $\phi$ is the distribution of the lengths of subsequent intersections between the considered phase and a randomly chosen line in direction $\phi$. For the estimation of chord length distributions from image data we refer to~\cite{ohser.2009}. Figure~\ref{fig:chordlength} shows the mean chord length distribution functions in $x$-, $y$- and $z$-direction for the eight samples to measure anisotropy effects. For all three phases, it can be seen that the chord length distribution functions in $x$- and $y$- directions are nearly identical and chord lengths in $x$- and $y$-direction are significantly larger than the ones in $z$-direction. The chord length distribution quantifies the elongation of phases in certain directions and is thus -- besides local percolation probabilities -- a further measure for the influence of uniaxial compression onto the microstructure.

\subsection{Results and discussion}

The presented statistical analysis characterizes the microstructure of tablets based on 3D images with a voxel size of $0.438~\upmu\mathrm{m}$. Thus, pores with a diameter below the resolution threshold are not taken into account in the analysis, which leads to (slightly) different values of porosity and specific surface area compared to experimental results. Being aware that the analysis is performed on a certain length scale, which is determined by the voxel resolution, statistical image analysis nevertheless allows for the computation of microstructure characteristics, which are not accessible by experiments. The performed analysis shows that the interface area between pores and MCC, as well as between pores and API grows linearly in porosity for the considered material. The computation of local percolation probabilities and chord length distribution functions allows for a quantitative analysis of the impact of compaction on the
microstructure, which can complement empirical models such as on the prediction of compactibility, compressibility and tabletability of compacts~\citep{imbert.1997}.

A promising and powerful application of the presented trinarization, which identifies the different components, consists of a further discretization which allows for numerical modeling of multicomponent formulations by means of the discrete element method (DEM). In particular, breakage and propagation of cracks can be numerically modeled. In future work, the statistically representative volume element size with respect to mechanical properties will be determined based on DEM simulations in the spirit of~\cite{dirrenberger.2014}.

In order to reinforce the current results, it might be necessary to include further microstructure characteristics in order to describe the microstructure influence on mechanical properties. Beyond local percolation probabilities, which are considered in the present paper as connectivity property, the quality of connecting pathways, important for effective conductivity in two-phase microstructures and reflected in the notions of mean geodesic tortuosity and constrictivity~\citep{stenzel.2017}, could also affect the mechanical properties significantly.

%% file: conclusions.tex
A novel methodology to investigate the microstructure of Ibuprofen tablets based on 3D image data from synchrotron tomography has been presented. To be more precise, a new trinarization algorithm is developed, which allowed for the identification and labelling of the three constituting phases, i.e., pores, API and MCC. Here an algorithm using methods from statistical learning, namely random forests, is combined with a watershed-based algorithm. In contrast to using just one of these two algorithms, the presented hybrid approach matches the required challenges in the sense that it leads to a physically realistic trinarization. This means that the trinarization algorithm has not detected tiny constitutive parts of API within MCC particles and, on the other hand, long pores at the boundary of MCC particles are correctly detected. The trinarization allowed for an investigation of the microstructure by means of statistical image analysis, which is illustrated by the example of a given tablet. The results obtained from statistical image analysis showed a slight discrepancy with experimental measurements, due to the pores with diameter below the resolution threshold. Furthermore, image analysis showed that the volume fractions of the three phases vary up to 5~\% when considering 8 different cubic cutouts of the tablet with an edge length of 0.3~mm. Moreover, the computation of local percolation probabilities and chord length distribution functions enabled us to quantify the influence of compaction of the tablet on its microstructure. The presented methodology can be used to investigate the relationship between production parameters of the tablet, as the ratio of volume fractions of API and MCC or the pressure of compaction, and the microstructure characteristics, which influence, in turn, effective properties of the tablet as its strength and solubility kinetics. As a possible subject of future research, a further powerful outcome of the trinarization might be a numerical modeling of multicomponent formulations by DEM, in order to elucidate the microstructure influence on mechanical properties of the tablet.

%% file: figures0.tex
\begin{figure}[h]
	\centering
	\includegraphics[width = 0.45 \textwidth]{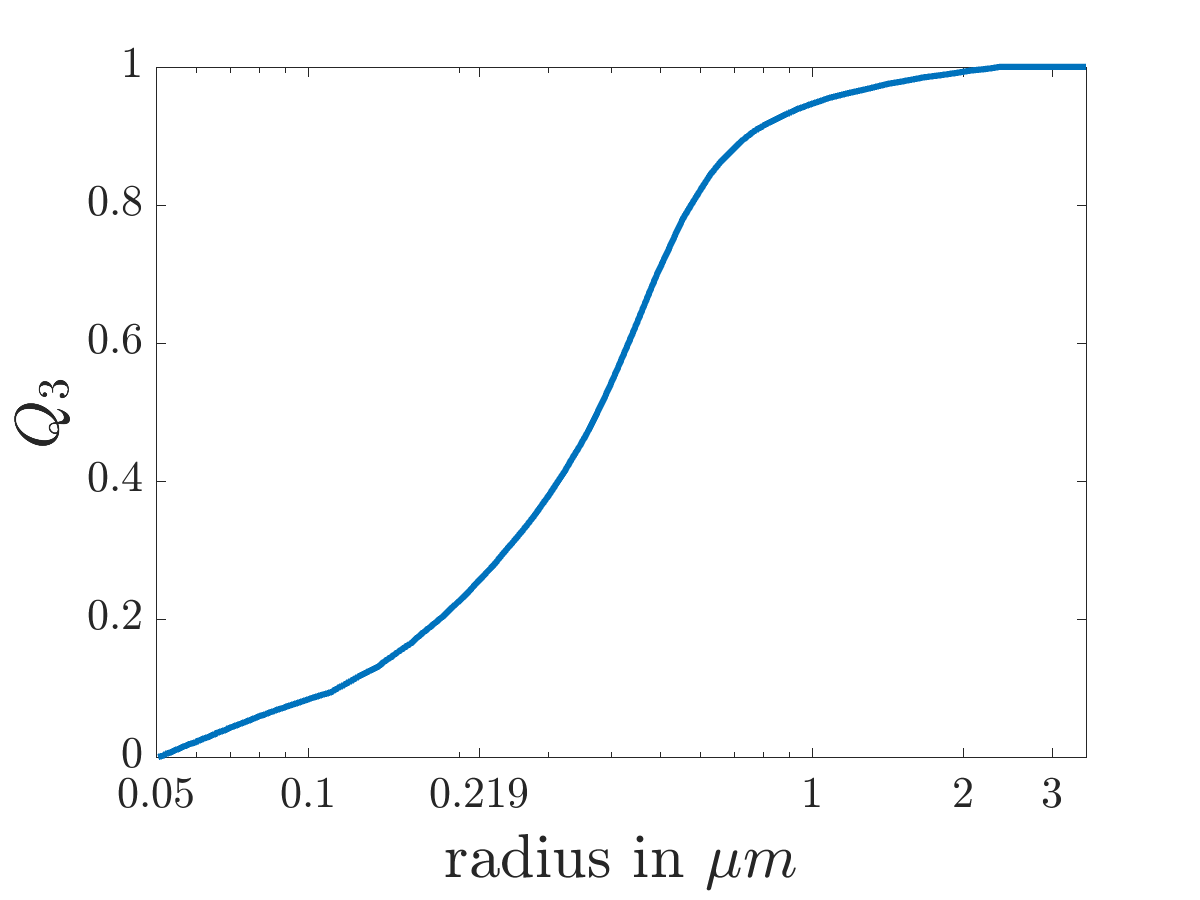}
	\caption{Experimentally determined MIP pore size distribution function $Q_3$ on a logarithmic scale.}\label{fig:Q3}
\end{figure}

\newpage

\begin{figure}[h]
	\centering
	\includegraphics[width = \textwidth]{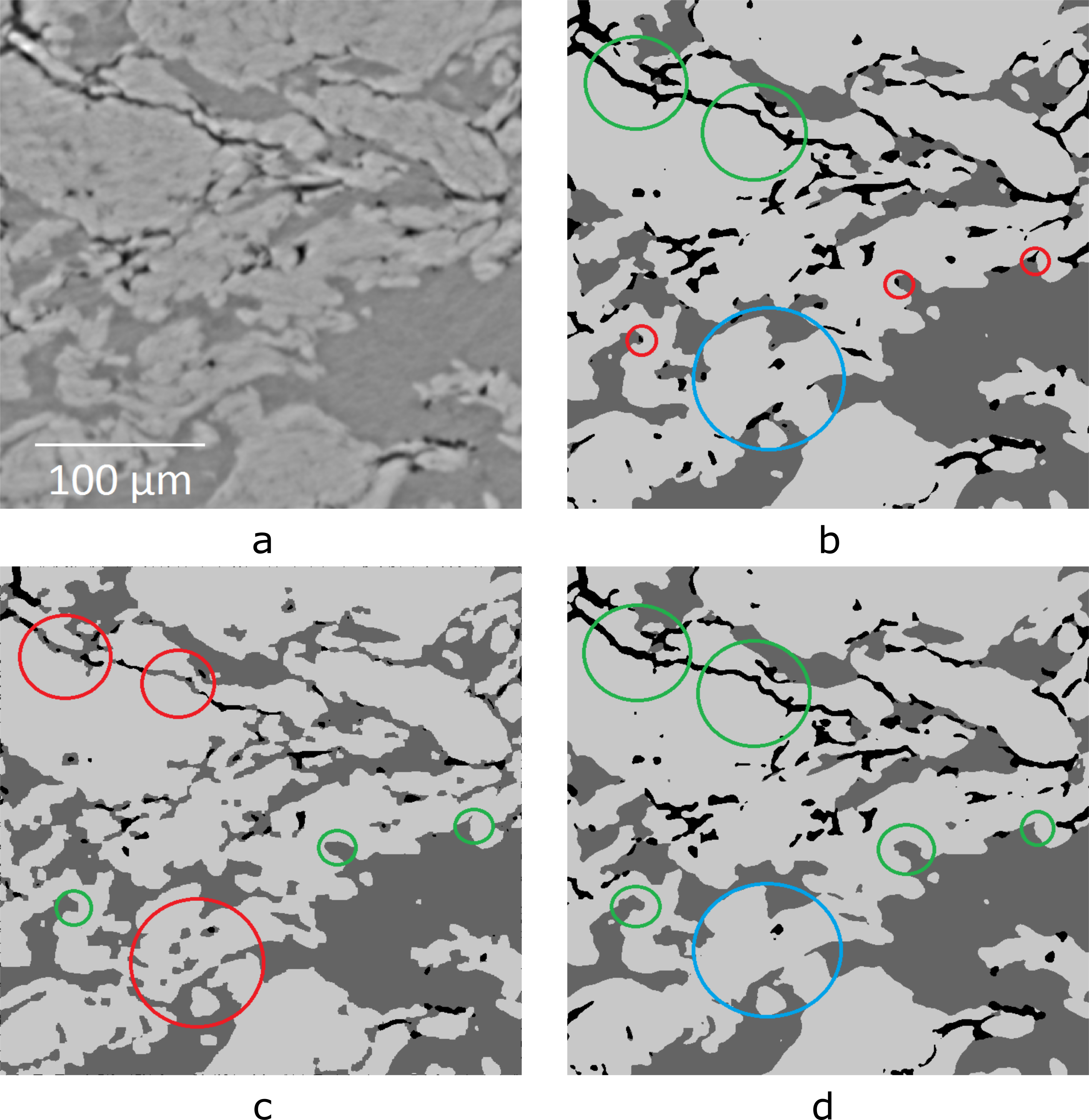}
	\caption{2D slice of a cutout of greyscale image data in the $xz$-plane (a) and different types of trinarization, i.e. the trinarization by the random forest algorithm (b), the watershed-based trinarization (c) and the hybrid approach (d). After trinarization, pores are visualized in black, API in dark grey and MCC in light grey. Examples for regions, where the corresponding trinarization algorithm works well are encircled in green and blue. Regions, where the algorithm leads to misclassifications, are in encircled in red.}\label{fig:tri2D}
\end{figure}

\newpage

\begin{figure}[h]
	\centering
	\includegraphics[width = \textwidth]{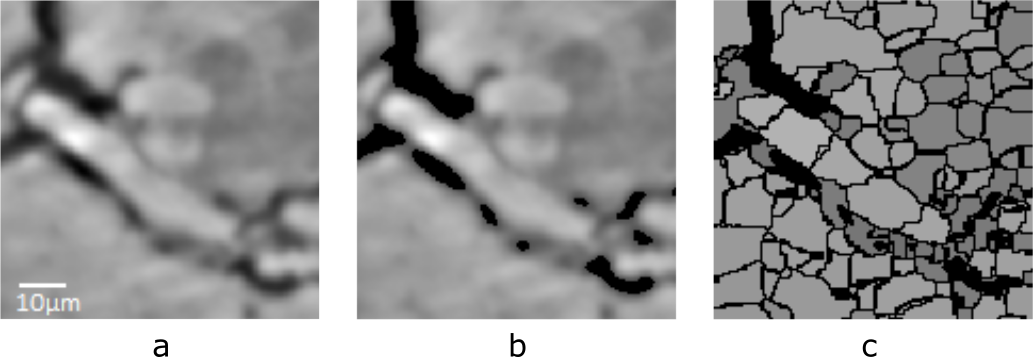}
	\caption{Visualization of the watershed segmentation at the example of a cutout of the same 2D slice, which is considered in Figure 2. At first the pores are determined (b, pores are represented in black) from the original greyscale image (a). Then, the image is partitioned in different watershed basins (c). The boundaries of watershed basins are represented in black and all voxels of each watershed basin are labeled with the average greyscale value of the basin.}\label{fig:watershed2D}
\end{figure}

\newpage

\begin{figure}[h]
	\centering
	\includegraphics[width = \textwidth]{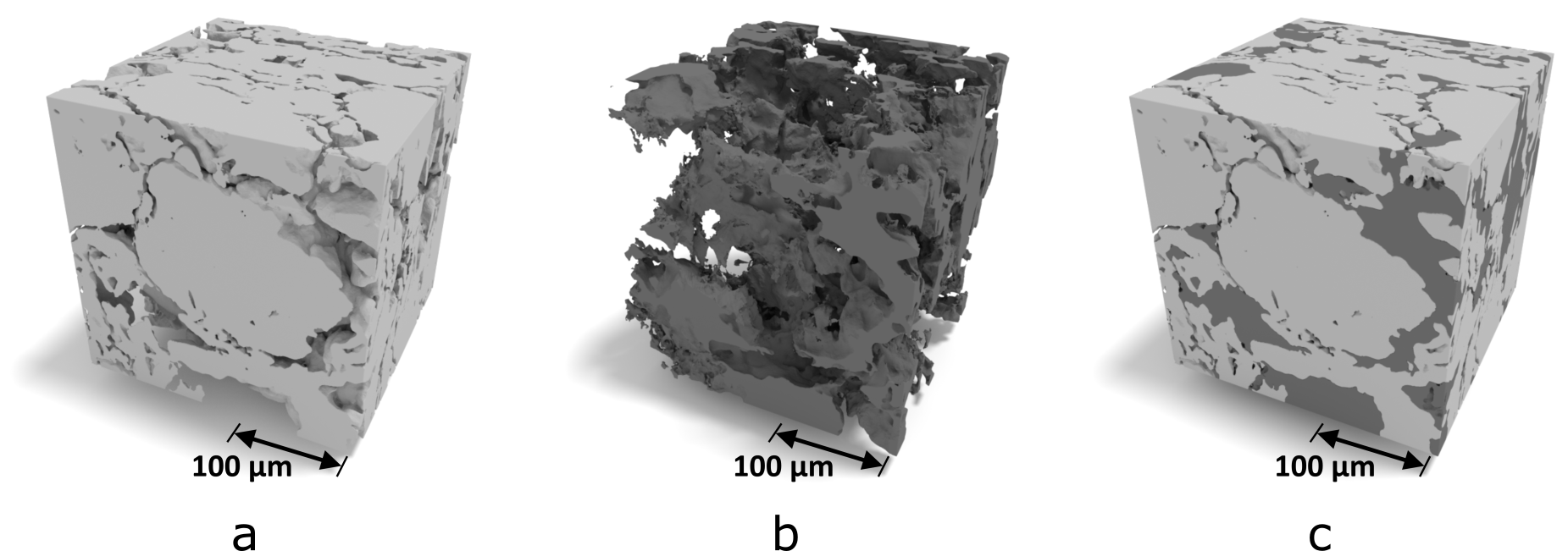}
	\caption{3D visualization of MCC (a), API (b) and both components (c) determined by the hybrid approach combining a  random forest algorithm with a watershed-based trinarization, where a cube of side length 306.6 $\upmu$m is used for visualization.}\label{fig:3Dvis}
\end{figure}

\newpage

\begin{figure}[h]
	\centering
	\includegraphics[width = \textwidth]{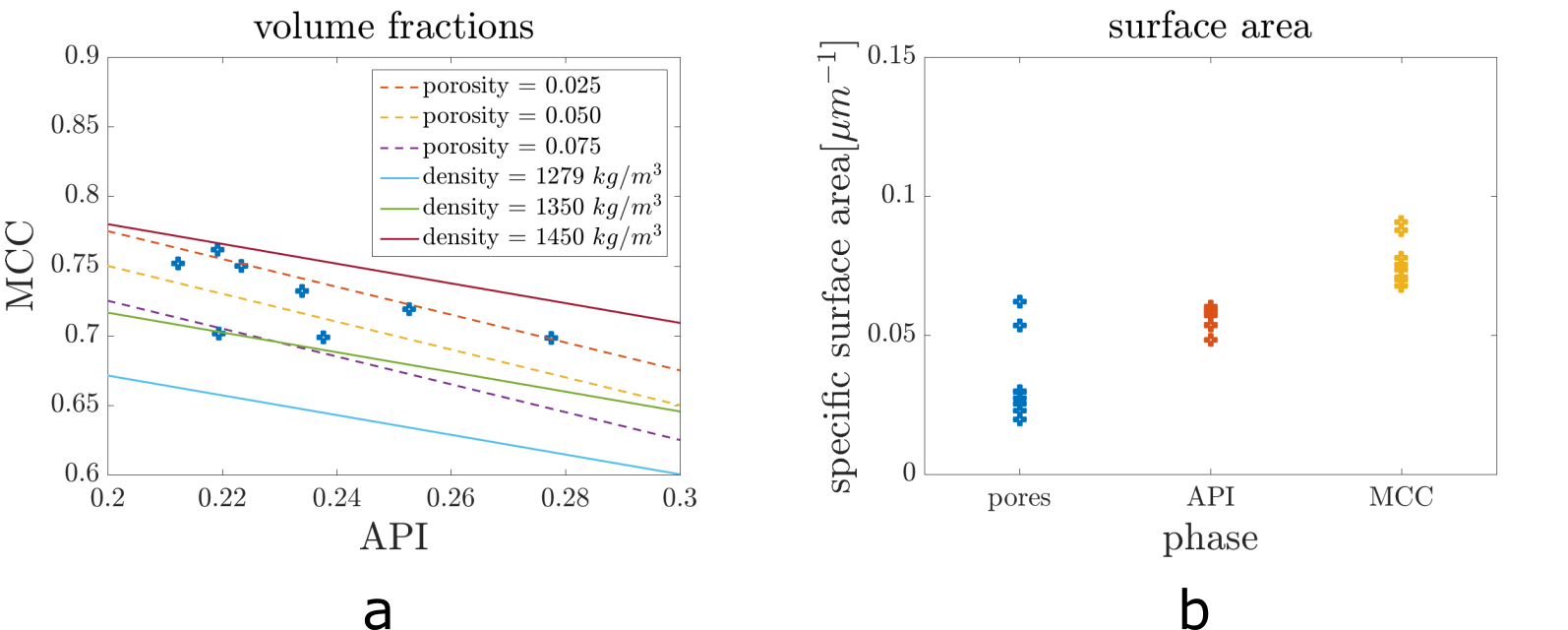}
	\caption{The volume fractions of API and MCC are represented by a blue dot for each of the eight cubic cutouts (a). Porosity as well as density of the cutouts are uniquely determined by volume fractions of API and MCC. The dashed/straight lines show volume fractions of MCC over the volume fractions of API for given values of porosity/density. Specific surface areas of the three phases (b).}\label{fig:vol_surf}
\end{figure}

\newpage

\begin{figure}[h]
	\centering
	\includegraphics[width = \textwidth]{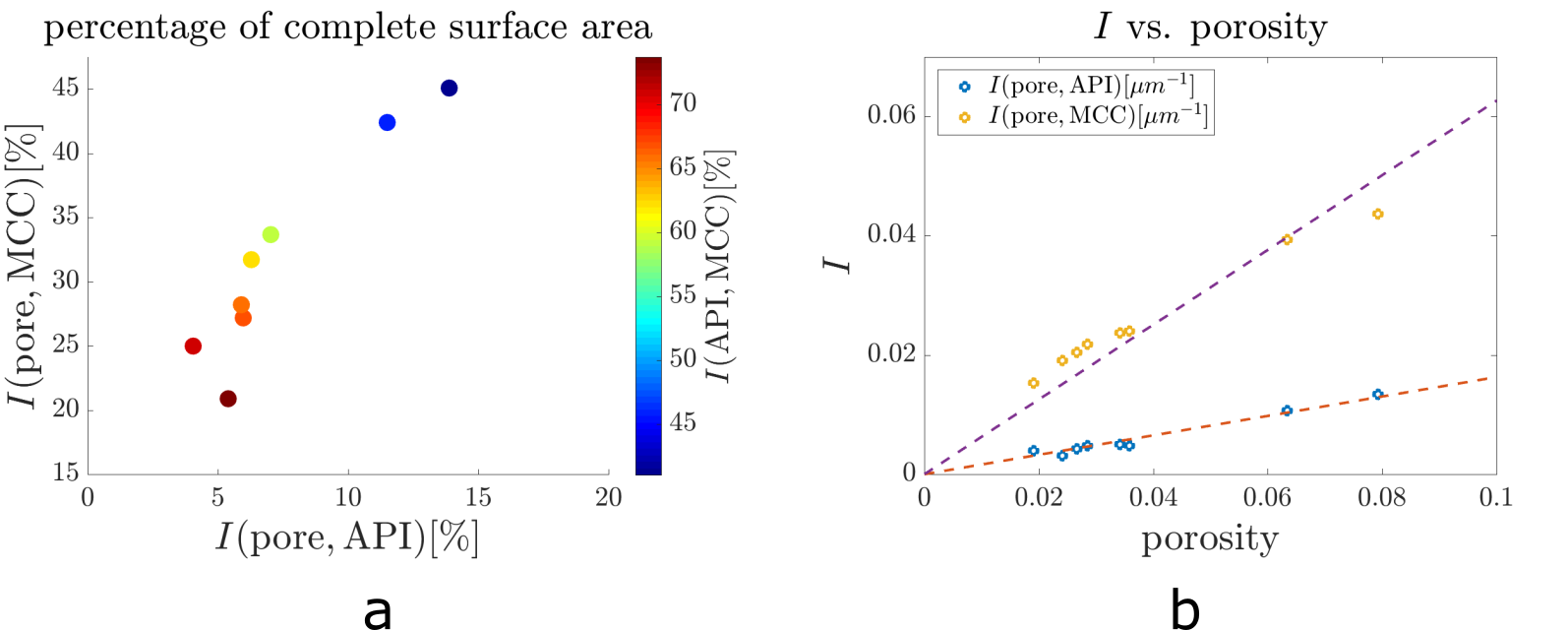}
	\caption{Surface areas of interfaces per unit cube are visualized: For the different interface areas per unit cube, their percentage of total surface area is visualized. The values of $I(\mathrm{Pore},\mathrm{MCC})$ are plotted over $I(\mathrm{Pore},\mathrm{API}),$ while the value of $I(\mathrm{API},\mathrm{MCC})$ is indicated by the color bar  (a). The values of $I(\mathrm{Pore},\mathrm{API})$ and $I(\mathrm{Pore},\mathrm{MCC})$ correlate strongly with porosity (b).}\label{fig:interface}
\end{figure}

\newpage

\begin{figure}[h]
	\centering
	\includegraphics[width = \textwidth]{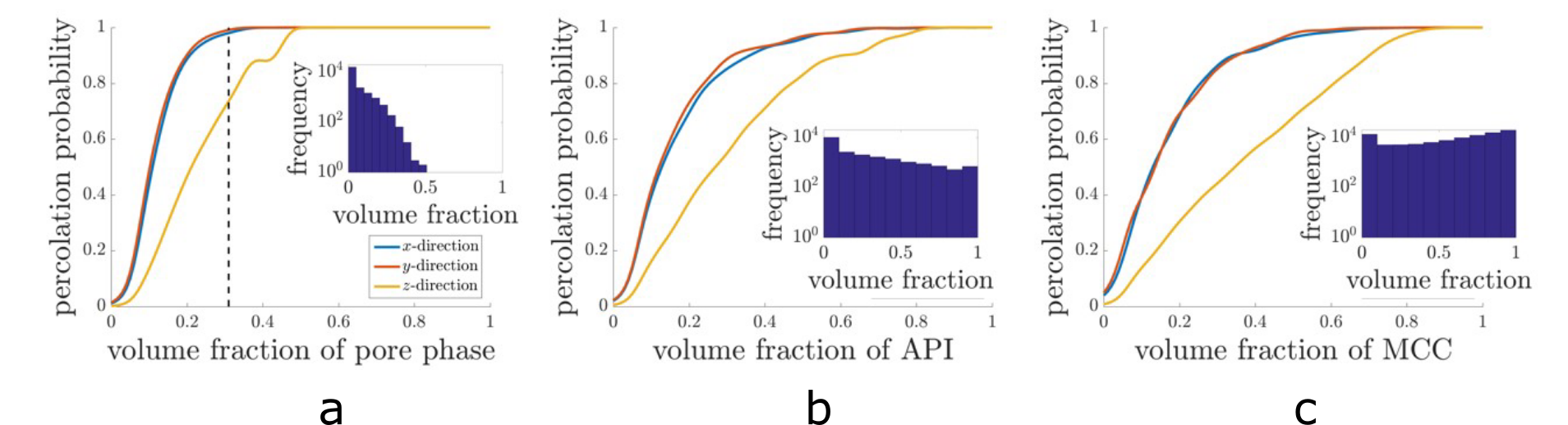}
	\caption{Local percolation probabilities estimated for pore phase (a), API (b) and MCC (c) based on sub-cubes with an edge length of 22~$\upmu$m. Percolation probabilities are computed for $x$-, $y$-, and $z$-direction. The blue histograms show the frequency of volume fractions (of the corresponding phase) within the considered sub-cubes. For the estimation of local percolation probabilities of the pore space, we consider the values of $P(v)$ to be meaningful, which are located on the left side of the black dotted line.}\label{fig:percolation}
\end{figure}

\newpage

\begin{figure}[h]
	\centering
	\includegraphics[width = \textwidth]{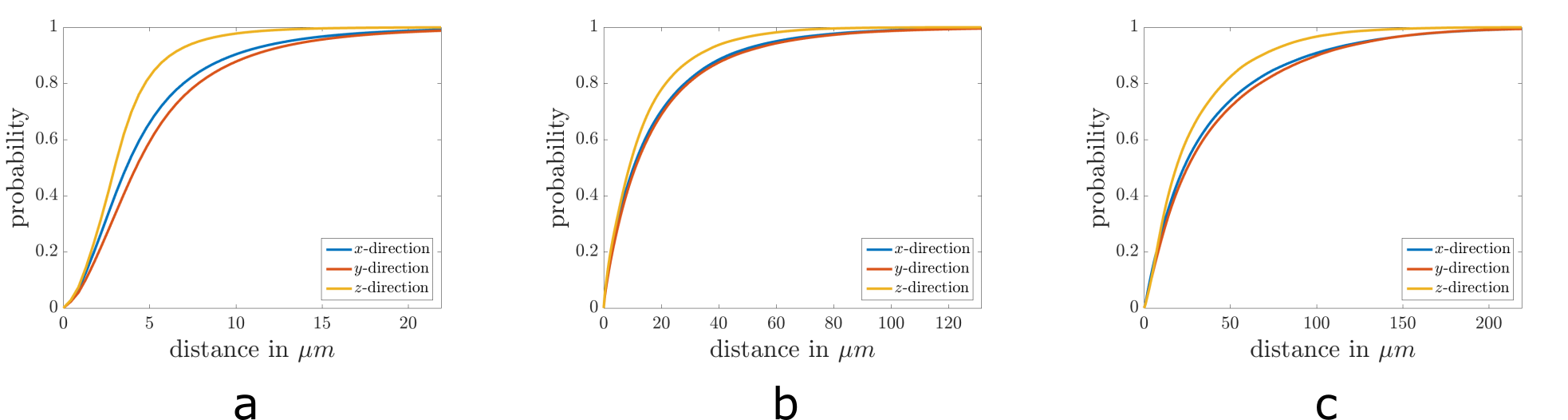}
	\caption{Mean chord length distribution functions estimated from image data. Chord lengths are computed for pores (a), API (b) and MCC (c).}\label{fig:chordlength}
\end{figure}

%% file: tables.tex
\begin{table}[h]
	\centering
	\caption{Physical properties of the powders of interest: quantiles of volumetric/mass cumulative PSD and true/skeletal density.}\label{tab:exp_results}
	\begin{tabular}{lccccc}
		Material & $X_{10} [\upmu\mathrm{m}]$ & $X_{50} [\upmu\mathrm{m}]$ & $X_{90} [\upmu\mathrm{m}]$ & Span $[-]$ & True Density [$\mathrm{kg} \cdot \mathrm{m}^{-3}$]   \\ \hline
		MCC & 43.9 & 161.5 & 320.1 & 1.71 & 1573.0 \\ \hline
		Ibuprofen & 7.7 & 25.4 & 69.2 & 2.42 & 1115.3
	\end{tabular}
\end{table}